\documentclass[12pt]{iopart}

\usepackage{graphicx}
\usepackage{booktabs}
\usepackage{paralist}
\usepackage{amsfonts,amssymb}
\usepackage{bm}
\usepackage{bbm}

\begin{document}

\title[Adjustable self-loop on discrete-time quantum walk and its application in spatial search]{Adjustable self-loop on discrete-time quantum walk and its application in spatial search}

\author{Wang Huiquan$^{*}$, Zhou Jie, Wu Junjie$^{\dag}$, Yi Xun}

\address{State Key Laboratory of High Performance Computing, National University of Defense Technology\\
College of Computer, National University of Defense Technology\\
Changsha, 410073, People's Republic of China}
\ead{$^{*}$huiquanwang@quanta.org.cn, $^{\dag}$junjiewu@nudt.edu.cn}
\vspace{10pt}

\begin{abstract}
How self-loops on vertices affect quantum walks is an interesting issue, and self-loops play important roles in quantum walk based algorithms. However, the original model that adjusting the effect of self-loops by changing their number has limitations. For example, the effect of self-loops cannot be adjusted continuously, for their number must be an integer. In this paper, we proposed a model of adjustable self-loop on discrete-time quantum walk, whose weight is controlled by a real parameter in the coin operator. The proposed method not only generalises the situations where the number of self-loops is an integer, but also provides a way to adjust the weight of the self-loop continuously. It enhances the potential of self-loops in applications. For example, we improve the success rate of the quantum walk based search on a $20\times20$ two-dimension lattice from $23.6\%$ to $97.2\%$ by the proposed method. And the success rate of the improved search, which only scales as $O(1/\log{N})$ before being improved, even increases slightly with the size of the lattice. Besides two-dimension lattices, we find that this method is also effective on three-dimension lattices. To the best of our knowledge, this is the first time that such an improvement is achieved on the quantum walk based spatial search.
\end{abstract}

%
%
%
%
%

\section{Introduction}

Quantum walk is the counterpart of classical random walk for the particle whose motion is subject to quantum mechanics\cite{Aharonov1993Quantum}. It can be viewed as a quantum particle spreading on an undirected graph. Quantum interference from multiple walk paths makes the quantum walk has many novel and interesting features\cite{Ashwin2000Quantum,Knight2003Quantum,Antoni2004Quasiperiodic}. As its classical counterpart, the quantum walk is also widely recognized as a powerful tool to develop algorithms. Based on quantum walks, many quantum algorithms have been proposed, such as the quantum walk based search\cite{shenvi2003quantum,lovett2012spatial,Portugal2013Quantum}, exponentially faster hitting\cite{Childs2002Exponential,Kempe2003Discrete,ANDRIS2004Quantum} and graph isomorphism algorithms\cite{douglas2008classical,emms2009graph1,emms2009graph2,qiang2012enhanced,Wang2015A}.

In the model of quantum walks, topology of the graph affects the evolution of the walker greatly, so using special graph structure to apply force to quantum walks is a commonly used strategy to solve computational problems. Such as graph isomorphism algorithms by Emms\cite{emms2009graph1,emms2009graph2}, and the universal quantum computation schemas by Childs\cite{Childs2009Universal,Childs2013Universal} and Lovett\cite{Lovett2009Universal}. In the works by Emms\cite{emms2009graph1,emms2009graph2}, a special auxiliary graph constructed by two similar graphs which to be matched is used to evolve quantum walks, then the matching result can be reflected by the final state of the walker. In the work by Childs\cite{Childs2009Universal,Childs2013Universal} and Lovett\cite{Lovett2009Universal}, some specially designed graphs are used to evolve quantum walks, to realise certain quantum logical operations on the walker's state.

In this paper, we focus on a basic structure to apply force to quantum walks, i.e. the self-loop, and its application. We consider the self-loop on the model of discrete-time quantum walk (DTQW)\cite{Aharonov2001Quantum} only. How the self-loop affects one dimensional DTQW has been studied\cite{Norio2005One,Falkner2014Weak,S2014Limit,Li2015One}. And in 2015, TG Wong gives an example to apply the self-loops to improve the DTQW based search on the complete graph\cite{Wong2015Grover}. In TG Wong's work, he proves that if we attach one self-loop to each vertex on the complete graph and then perform the DTQW based search, its success rate can be improved from 50\% to 100\%, but more self-loops will make the success rate decreases.

Considering Wong's results, a natural question is whether his idea can be extended to improve the DTQW based search on other structures, such as the two-dimensional lattice, on which the success rate of DTQW based search only scales as $O(1/\log N)$\cite{lovett2012spatial}, where $N$ is the size of the lattice. However, it fails. We cannot improve the success rate of DTQW based search on two-dimensional lattices, no matter how many self-loops we attach on each vertex. In next section, we will introduce the result of our test.
Through the test, we find that the self-loop has a huge effect on the performance of DTQW based search on two-dimensional lattices, but the original model that adjusts the strength of self-loops' effect on DTQW by changing self-loops' number has two limitations:
\begin{itemize}
  \item In that method, the strength of self-loops' effect can only be adjusted discretely, because the number of self-loop must be an integer. So the effect of self-loops which between the effects of $n$ self-loops and $n+1$ self-loops cannot be tested by the original model, where $n$ is an arbitrary non-negative integer.
  \item When we attach self-loops on the graph, the dimension of the walker's state space increases with the number of self-loops. It may bring inconvenience in both simulation and physical implementation.
\end{itemize}

To break the limitations above, we propose a new model to adjust the strength of self-loops' effect on DTQW. In the proposed method, we attach only one self-loop on each vertex of the graph, and adjust the effect of this self-loop with a specially designed coin operator on DTQW, in which there is a real parameter $n$ to control the weight of the self-loop. When $n$ is zero, the self-loop do not affect the quantum walk, just as there is no self-loop in the graph; when $n$ is a positive integer, it is equivalent to attaching $n$ self-loops on each vertex in the standard DTQW. So the parameter $n$ is equivalent to the number of self-loops in the standard DTQW when it is a non-negative integer. However, by setting the parameter $n$ to a non-integer, we are able to explore the situations where the effect of self-loops is between the effects of $\lfloor n\rfloor $ self-loops and $\lfloor n\rfloor+1$ self-loops, where $\lfloor n\rfloor $ is largest integer not greater than $n$.

This model will extend the application of the self-loop on DTQW based algorithms. Through the proposed method, we study the effect of the self-loop on the DTQW based search on two-dimension lattices again. We find that if we set the weight of each adjustable self-loop equal to the vertex's degree centrality (which is a non-integer) when perform the DTQW based search on two-dimension lattices, the success rate of the search, which only scales as $O(1/\log{N})$ originally, can always be improved to nearly 100\%, and it even increases slightly with the size of the lattice. Then we verify that this conclusion can be generalised to the DTQW based search on three-dimension lattices. To the best of our knowledge, this is the first time that such an improvement is achieved on the DTQW based search on two-dimension lattices and three-dimension lattices.

The paper is organized as follows. Section 2 provides an introduction to discrete-time quantum walk, and some background about the work by TG Wong\cite{Wong2015Grover} and our previous test. Section 3 describes our proposed method. Section 4 applies the proposed method to improve the DTQW based search on two-dimension lattices and three-dimension lattices. Conclusions are presented in Section 5.

\section{Background}

Discrete-time quantum walk (DTQW) is a quantization of the classical random walk.
In the classical case, we often suppose a walker starts from one vertex of an undirected graph, and moves to any one of adjacent vertices with equal probability at each step. The standard DTQW is a quantization of this process. In DTQW, the walker's position is denoted by a vector in a Hilbert space $\mathcal{H}_s$, whose computational basis are $\{|j\rangle_s:j \in \{1,2,...N\}\}$, where $N$ is the number of vertices in the graph. To describe the direction of the walker, we attach a coin space $\mathcal{H}_c$ on every vertex, whose computational basis are $\{|k\rangle_c:k \in \{1,2,...N\}\}$, which means the position of the walker at next step. Then if we perform the DTQW on a graph with $N$ vertices, we use the Hilbert space $\mathcal{H} = \mathcal{H}_s \otimes \mathcal{H}_c = span\{|j\rangle_s\otimes|k\rangle_c:i,j\in \{1,2,...,N\}\}$ as the space where the walker evolves. The basis $|j\rangle_s\otimes|k\rangle_c$ means that the walker is at the vertex $v_j$ and it will move to the vertex $v_k$ at next step. In this paper, we often express the $|j\rangle_s\otimes|k\rangle_c$ as $|j\rangle_s|k\rangle_c$ or $|j,k\rangle$ for simplicity. At each step of the DTQW, we act on the state of the walker with a coin operator followed by a shift operator. The coin operator splits the walker on each vertex into a quantum superposition of different directions, and then the shift operator moves the walker to the next vertex based on its direction. We often act a \emph{Grover operator} on the \emph{coin space} of each vertex as the coin operation\cite{Portugal2013Quantum}. The Grover operator on the vertex $v_j$ is defined as
\begin{equation}\label{E_Grover_operator}
  G_j = 2|D_j\rangle \langle D_j|-I
\end{equation}
where $|D_j\rangle$ is the diagonal state\cite{Portugal2013Quantum} in the coin space of $v_j$, i.e. the equal superposition of each direction of the walker on $v_j$. For example, if vertex $v_j$ has $m$ neighbors and has no self-loop, the state $|D_j\rangle$ for this vertex is
\begin{equation}\label{diagonal state}
  |D_j\rangle = \frac{1}{\sqrt{m}}\sum\limits_{k \in B_j}{|k\rangle}_c
\end{equation}
where $B_j$ is the set of neighbors of vertex $v_j$. The state $|D_j\rangle$ for the vertex has self-loops will be introduced in the next section. The coin operator of DTQW is defined as
\begin{equation}\label{coin_operator_std}
  C = \sum\nolimits_j |j\rangle_s\langle j|\otimes G_j
\end{equation}
The shift operator moves the walker to the next position (vertex) according to its coin state, and this operator is defined as
\begin{equation}\label{step_operator}
  S = \sum\limits_{j,k = 1}^N {|j\rangle_s |k\rangle_c\ \langle k|_s \langle j|_c}= \sum\limits_{j,k = 1}^N {\left| {j,k} \right\rangle \left\langle {k,j} \right|}
\end{equation}
Then the evolution of the DTQW on an undirected graph can be expressed as
\begin{equation}\label{evolution_of_DTQW}
  \left| {{\psi_t}} \right\rangle  ={(SC)^t}\left| {{\psi_0}} \right\rangle = {U^t}\left| {{\psi_0}} \right\rangle
\end{equation}
where $U=SC$, $\left| {{\psi_t}} \right\rangle$ is the final state of the DTQW after $t$ steps, and $\left| {{\psi_0}} \right\rangle$ is the initial state.

An important application of DTQW is the DTQW based search algorithm. The method to develop a search algorithm with DTQW is marking vertices in the graph by a coin operator which distinguishes between target and non-target vertices. The phase-flip coin is the most common coin operator in DTQW based search:
\begin{equation}\label{Search_coin_operator}
  C^{flip} = \sum\nolimits_j |j\rangle_s\langle j|\otimes (-1)^{f(j)}G_j
\end{equation}
where the function $f$ is the querying function:
\begin{equation}\label{querying_function}
  f(x)=\left\{\begin{array}{ll}
                1, & if\ x\ is\ a\ target\ for\ the\ search \\
                0, & otherwise
              \end{array}\right.
\end{equation}
Then if we set the initial state of walker as $|\psi_0\rangle = \sum_{j=0}^N{\sum_{(j,k)\in E}{\frac{1}{\sqrt{N\cdot Deg(j)}}}}|j,k\rangle$, where $Deg(j)$ means the degree of $v_j$, and perform the DTQW by the evolution operator $U=S\cdot C^{flip}$, the probability of the walker will concentrate to the target vertex after certain steps. We evaluate the success rate of DTQW based search by the first peak of probability on the target vertex, and evaluate the computational complexity by how the step to the first peak scales with the size of the graph. More detail about the DTQW based search can be found in references\cite{lovett2012spatial,Portugal2013Quantum}.

\begin{figure}[htbp]
  \centering
  \includegraphics[width=1.0\textwidth]{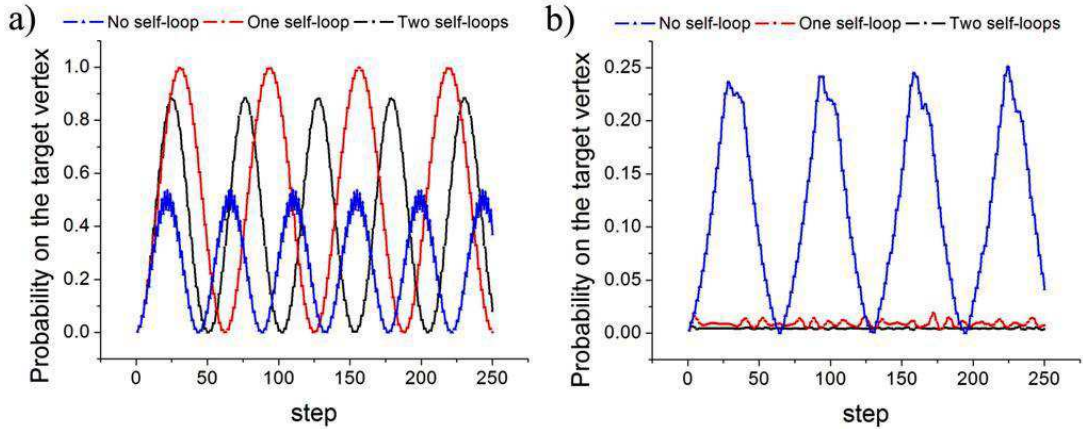}\\
  \caption{Performance of the quantum walk based search on graphs without self-loop, and with one and two self-loops respectively. The figures record how probabilities on the target vertex vary with steps, and the success rate of DTQW based search is reflected by the first peak of the probability\cite{lovett2012spatial,Wong2015Grover}. a) The DTQW based search on the complete graph with $400$ vertices. The success rate of DTQW based search on the complete graph without self-loop is $53\%$. If we attach one self-loop on each vertex, the success rate is enhanced to $100\%$. While if we attach two self-loops on each vertex, the success rate decrease to $88\%$. b) The DTQW based search on a $20\times20$ two-dimensional lattice. The success rate of DTQW based search on the lattice without self-loop is $23.6\%$. However, if we attach one self-loop or two self-loops on each vertex, the success rate decreases to no more than $2\%$.}\label{Complete_Mesh_loop}
\end{figure}

A variety of works have been done to investigate characteristics of DTQW based search on various structures, such as on complete graphs\cite{Reitzner2009Quantum,Wong2015Grover}, two-dimensional lattices\cite{lovett2012spatial} and complex networks\cite{Berry2010Quantum}. The most closely related work to this paper is the work by TG Wong\cite{Wong2015Grover} and NB Lovett\cite{lovett2012spatial}. In 2015, TG Wong investigates the quantum walk search on the complete graph, and proves that the self-loop affects the success rate of DTQW based search on complete graphs greatly. If the DTQW based search is performed on the complete graph without self-loop, the success rate is only about 50\%. But if we attach one self-loop on every vertex, the success rate of the search is enhanced to 100\%, and if we add more self-loops, the success rate will decrease again. Figure \ref{Complete_Mesh_loop}(a) shows the performance of DTQW based search algorithm on the complete graph without self-loop, and with one and two self-loops respectively, where the complete graph has 400 vertices.
Another work which we care about is the investigation by NB Lovett, which shows that the success rate of DTQW based search on two-dimensional lattice scales as $O(1/\log_2 N)$ only\cite{lovett2012spatial}, where $N$ is the number of vertices in the lattice. We perform the DTQW based search on a $20\times20$ two-dimensional lattice without self-loop for example, and the result is shown in Figure \ref{Complete_Mesh_loop}(b). We can see that the success rate of the DTQW based search in this test is only about $23\%$. Then a natural question is whether we can enhance the success rate by attaching self-loops on each vertex of the two-dimensional lattice, as the method in the work by TG Wong\cite{Wong2015Grover}. However, it fails. Figure \ref{Complete_Mesh_loop}(b) shows how attaching self-loops affect the DTQW based search on the two-dimensional lattice with 400 vertices. In this investigation, we find two limitations on the original method. Firstly, the effect of self-loops can only be adjusted discretely, for the number of self-loop must be an integer. Secondly, the more self-loops we attach on the vertex, the bigger the state space is. It will take more resources in both stimulation and physical implementation.

In this paper, we proposed a method to break these two limitations. More specifically, we attach only one self-loop on each vertex, and do not change the structure of the graph again. Then we propose a special coin operator to adjust the weight of the self-loop on DTQW. The detail is as follows.

\section{Adjustable self-loop on discrete-time quantum walk}

\begin{figure}[htbp]
  \centering
  \includegraphics[width=0.9\textwidth]{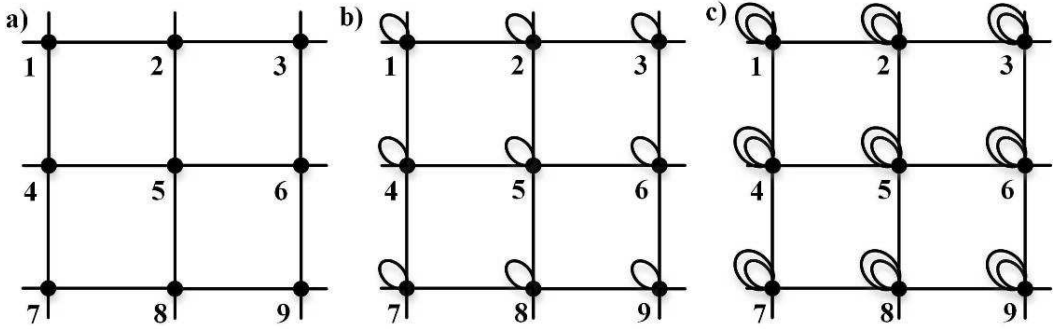}\\
  \caption{Examples of attaching self-loops on the graph. a) A two-dimension lattice without self-loop. b) A two-dimension lattice with one self-loop on each vertex. c) A two-dimension lattice with two self-loops on each vertex.}\label{graph_loop}
\end{figure}

In the previous works, the weight of self-loops is controlled by their number, as figure \ref{graph_loop} shows. By that method, the weight of self-loops cannot be adjusted continuously, for the number of self-loops must be an integer.
Let us observe how attaching self-loops affect the evolution of DTQW theoretically. Suppose $v_j$ is a vertex without self-loop and it has $m$ neighbors. The Grover operator on $v_j$ is defined as $G_j = 2|D_j\rangle \langle D_j|-I$ where $|D_j\rangle$ is the diagonal state as equation (\ref{diagonal state}). If we attach $n$ self-loops on $v_j$, the additional self-loops make the diagonal state (equation (\ref{diagonal state})) change to
\begin{equation}\label{E_Diagonal_state_n_loop}
  |D_j\rangle = \frac{1}{\sqrt{m+n}}\left(\sum\limits_{k \in B_j}|k\rangle_c + \sum_{i=1}^n|j^{(i)}\rangle_c\right)
\end{equation}
where $B_j$ is the set of neighbors of $v_j$, and $|j^{(i)}\rangle$ means the base of the $i$th self-loop of $v_j$. We can see that the self-loops affect the DTQW by changing the coin operation of the walker. More specifically, the more self-loops, the more probability of the walker stay at the current position. From equation (\ref{E_Diagonal_state_n_loop}) we also see that attaching $n$ self-loops make the dimension of the Grover operator (equation (\ref{E_Grover_operator})) increase from $m$ to $m+n$.

In this section, we propose a method to replace the multiple self-loops with one adjustable self-loop whose weight on DTQW can be adjusted by a new coin operator.
To realise this goal, we need to make a modification on the original Grover operator (equation (\ref{E_Grover_operator})). We propose a new coin operator as follows:
\begin{equation}\label{E_Grover_loop_operator}
  G_j'(n) = 2|D_j'(n)\rangle \langle D_j'(n)|-I
\end{equation}
where $n$ represents the weight of the self-loop, and $|D_j'(n)\rangle$ is a modification of the diagonal state in equation (\ref{E_Diagonal_state_n_loop}):
\begin{equation}\label{E_Diagonal_state_adjustable_loop}
  |D_j'(n)\rangle = \frac{1}{\sqrt{m+n}}\left(\sum\limits_{k \in B_j}|k\rangle_c + \sqrt{n}|j\rangle_c\right)
\end{equation}
In the original diagonal state, i.e. the equation (\ref{E_Diagonal_state_n_loop}), $n$ represent the number of self-loops on vertex $v_j$, and the dimension of $|D_j\rangle$ is affected by $n$ directly (the dimension is $m+n$, where $m$ is the number of neighbors of $v_j$). While in equation (\ref{E_Diagonal_state_adjustable_loop}), $n$ represent the weight of the only one self-loop on vertex $v_j$, and the dimension of $|D_j'(n)\rangle$ is constant (the dimension is $m+1$).
We call operator $G_j'(n)$ (equation (\ref{E_Grover_loop_operator})) as the \emph{Grover-loop} operator. Then we make a new coin operator for DTQW as follows.
\begin{equation}\label{coin_operator_loop}
  C'(n) = \sum\nolimits_j |j\rangle_s\langle j|\otimes G_j'(n)
\end{equation}
If we perform DTQW on the graph with one self-loop on each vertex (the graph in figure\ref{graph_loop}(b) for example) with the evolution operator $U=S\cdot C'(n)$, we can adjust the weight of the self-loop by the parameter $n$.

The proposed method has two characteristics as follows.
\begin{enumerate}
  \item If we perform DTQW by the evolution operator $U=S\cdot C'(n)$ and $n=0$, the self-loop has no effect on evolution of the walker, just as there is no self-loop on the graph.
  \item If we perform DTQW by evolution operator $U=S\cdot C'(n)$ and $n$ is an positive integer, it makes the distribution of the walker be equivalent to the situation where the evolution operator $U=S\cdot C$ (diagonal state is computed by equation(\ref{E_Diagonal_state_n_loop})) is used and there are $n$ self-loops on each vertex of graph.
\end{enumerate}
The first characteristic can be verified easily, because from equation (\ref{E_Grover_operator}) and equation (\ref{E_Grover_loop_operator}) we can verify that $G_j'(0)=G_j$.
Then we explain the second one as follows.
Without loss of generality, we use the graphs in figure \ref{graph_loop} for example. Suppose we perform DTQW on the graph in figure \ref{graph_loop}(b) by $U=S\cdot C'(2)$ and perform DTQW on the graph in figure \ref{graph_loop}(c) by $U=S\cdot C$. We use $\alpha_{jk}$ and $\beta_{jk}$ to represent the probability amplitudes on these two graphs respectively, and their subscript ($jk$) means they are probability amplitudes on the base $|j,k\rangle$. Suppose on the initial state, two walkers have the same distributions, which means that their probability amplitudes satisfy the condition as follows.
\begin{equation}\label{condition_equivalent}
   \left\{\begin{array}{ll}
      \alpha_{jk} = \beta_{jk} & k\neq j\\
      \alpha_{jk} = \sqrt{n}\beta_{jk}^{(i)} & k=j\ \mbox{(the probability amplitudes on self-loops)}
    \end{array}\right.
\end{equation}
where $n$ is the number of self-loops ($n=2$ in this example), and $i$ means the $i$th self-loop in the multiple self-loops. Since the self-loops on the same vertex are equivalent to each other, we suppose the probability amplitudes on them are equal. So $\beta_{jj}^{(1)} = \beta_{jj}^{(2)}$ in this example.
Then we will prove that the two walkers will always have the same distribution during DTQW, if they have the same distribution on the initial state. The main idea of the proof is to verify that both the coin operation and shift operation preserve the same distributions of the two walkers.
Suppose at a certain time, the walkers have the same distributions. Let us observe the evolution of probability amplitudes on vertex $v_5$ on each graph for example. The corresponding probability amplitudes are $\{\alpha_{52},\alpha_{54},\alpha_{56},\alpha_{58},\alpha_{55}\}$ and $\{\beta_{52},\beta_{54},\beta_{56},\beta_{58},\beta_{55}^{(1)},\beta_{55}^{(2)}\}$ respectively. The evolutions of these probability amplitudes after coin operation are shown as table \ref{T_Qiang_GI_add_edges}.
\begin{table}[htbp]
\caption{\label{T_Qiang_GI_add_edges}The evolution of probability amplitudes on vertex $v_5$ after coin operation. $\alpha_{jk}$ and $\beta_{jk}$ are the probability amplitudes before coin operation. $\alpha_{jk}'$ and $\beta_{jk}'$ are the probability amplitudes after coin operation.}
\begin{tabular*}{\textwidth}{@{}c*{15}{@{\extracolsep{0pt plus 12pt}}c}}
\br
One self-loop \& $U=SC'(2)$ & two self-loops \& $U=SC$\\
\mr
$\alpha_{52}'= \frac{\alpha_{52}+\alpha_{54}+\alpha_{56}+\alpha_{58}+\sqrt{2}\alpha_{55}}{6} - \alpha_{52}$ &  $\beta_{52}'= \frac{\beta_{52}+\beta_{54}+\beta_{56}+\beta_{58}+\beta_{55}^{(1)}+\beta_{55}^{(2)}}{6} - \beta_{52}$   \\
$\alpha_{54}'= \frac{\alpha_{52}+\alpha_{54}+\alpha_{56}+\alpha_{58}+\sqrt{2}\alpha_{55}}{6} - \alpha_{54}$ &  $\beta_{54}'= \frac{\beta_{52}+\beta_{54}+\beta_{56}+\beta_{58}+\beta_{55}^{(1)}+\beta_{55}^{(2)}}{6} - \beta_{54}$   \\
$\alpha_{56}'= \frac{\alpha_{52}+\alpha_{54}+\alpha_{56}+\alpha_{58}+\sqrt{2}\alpha_{55}}{6} - \alpha_{56}$ &  $\beta_{56}'= \frac{\beta_{52}+\beta_{54}+\beta_{56}+\beta_{58}+\beta_{55}^{(1)}+\beta_{55}^{(2)}}{6} - \beta_{56}$   \\
$\alpha_{58}'= \frac{\alpha_{52}+\alpha_{54}+\alpha_{56}+\alpha_{58}+\sqrt{2}\alpha_{55}}{6} - \alpha_{58}$ &  $\beta_{58}'= \frac{\beta_{52}+\beta_{54}+\beta_{56}+\beta_{58}+\beta_{55}^{(1)}+\beta_{55}^{(2)}}{6} - \beta_{58}$   \\
$\alpha_{55}'= \frac{\sqrt{2}(\alpha_{52}+\alpha_{54}+\alpha_{56}+\alpha_{58}+\sqrt{2}\alpha_{55})}{6} - \alpha_{55}$ &  $\beta_{55}'^{(i)}= \frac{\beta_{52}+\beta_{54}+\beta_{56}+\beta_{58}+\beta_{55}^{(1)}+\beta_{55}^{(2)}}{6} - \beta_{55}^{(i)}$   \\
\br
\end{tabular*}
\end{table}
From table \ref{T_Qiang_GI_add_edges} we can verify the relationships as equation (\ref{E_after_coin}).
\begin{equation}\label{E_after_coin}
    \begin{array}{c}
      \alpha_{52} = \beta_{52} \\
      \alpha_{54} = \beta_{54}  \\
      \alpha_{56} = \beta_{56}  \\
      \alpha_{58} = \beta_{58}  \\
      \alpha_{55} = \sqrt{2}\beta_{55}^{(i)}
    \end{array}\Longrightarrow
    \begin{array}{c}
      \alpha_{52}' = \beta_{52}' \\
      \alpha_{54}' = \beta_{54}'  \\
      \alpha_{56}' = \beta_{56}' \\
      \alpha_{58}' = \beta_{58}'  \\
      \alpha_{55}' = \sqrt{2}\beta_{55}'^{(i)}
    \end{array}
\end{equation}
where $\alpha_{jk}$ and $\beta_{jk}$ are the probability amplitudes before coin operation; $\alpha_{jk}'$ and $\beta_{jk}'$ are the probability amplitudes after coin operation.
So after coin operation, two walkers will still have the same distribution if they have the same distribution before. The shift operator only shifts the probability amplitudes according to the topology of the graph, and the probability amplitudes on the self-loop will not change after the shift operation. So the two walkers will also have the same distribution after the shift operation (the premise is that the two graphs have the same topology except self-loops). So we can conclude that both the coin operation and shift operation preserve the same distributions of the two walkers. These conclusions can be generalises to the situations where $n$ is an arbitrary non-negative integer. From the analysis above, we can see that if we use the operator $G_j'(n)$ on the vertex with one self-loop and $n$ is an non-negative integer (including zero), the effect will be the same as the situation where we attach $n$ self-loops on the vertex and then perform DTQW with the original Grover coin.

Besides the two characteristics above, more importantly, this method can be extended to the situations where the ``number'' of self-loops is \emph{not an integer}, just by setting the parameter $n$ in the operator $G_j'(n)$ to a non-integer.
\begin{figure}[htbp]
  \centering
  \includegraphics[width=0.65\textwidth]{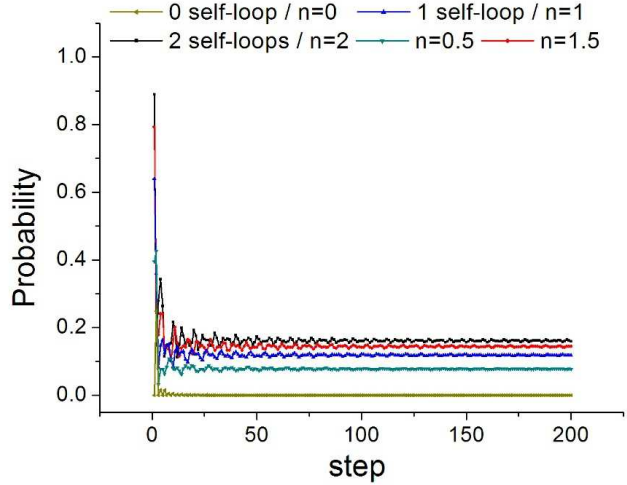}\\
  \caption{The comparison between the effects of adjusting the weight of self-loops by the new model and by the original model. The figure records how the probabilities on the start vertex ($v_c$) vary with step. The new method not only be equivalent to the original method when $n$ is non-negative integer (such as $n=0,1,2$), but also cover the situations where the ``number'' of self-loop is not an integer (such as $n=0.5,1.5$).}\label{compare_loops}
\end{figure}
For example, we perform the DTQW by $U'=SC'(n)$ on a $401\times401$ two-dimension lattice which has one self-loop on each vertex, and set $n$ as $0,0.5,1,1.5,2$ respectively. In these test, we suppose the walker start from the central vertex $v_c$ of the lattice, and set the initial state of DTQW as $|\psi_0\rangle = (|v_c,v_u\rangle+|v_c,v_r\rangle+|v_c,v_d\rangle+|v_c,v_l\rangle)/2$, where $v_u,v_r,v_d,v_l$ are vertices which above $v_c$, on the right side of $v_c$, under $v_c$ and on the left side of $v_c$ respectively (refer to figure \ref{graph_loop}). Then we record the probability of the walker on $v_c$ over 200 steps. The results are shown in figure \ref{compare_loops}. For comparison, we also perform the DTQW with the original Grover coin operator ($U=SC$) on the $401\times401$ two-dimension lattice without self-loop, and with one and two self-loops respectively. The results are also shown in figure \ref{compare_loops}.
From the results we can see that the new method not only be equivalent to the original method when $n$ is an integer ($n=0,1,2$), but also cover the situations where the ``number'' of self-loop is not an integer ($n=0.5,1.5$). The results show that we successfully realise the self-loop whose weight is between two successive integers by the proposed method.

In this paper, we will not explore the performance of this method on various structures further, although it may be interesting. Instead, we will give an example to show how this new method enhances the potential of DTQW in the application.

\section{The application of the proposed method in spatial search}

In this section, we use the discrete-time quantum walk (DTQW) based search as an example to show how the proposed method enhances the potential of DTQW in the application. In the section ``Background'' we have shown that attaching one self-loop on each vertex boosts the success rate of DTQW based search on the complete graph from about 50\% to 100\%, but fails to improve the DTQW based search on two-dimension lattices. We will show that, by the proposed adjustable self-loop, we are able to improve the success rate of DTQW based search on two-dimension lattices significantly.

\begin{figure}[htbp]
  \centering
  \includegraphics[width=0.99\textwidth]{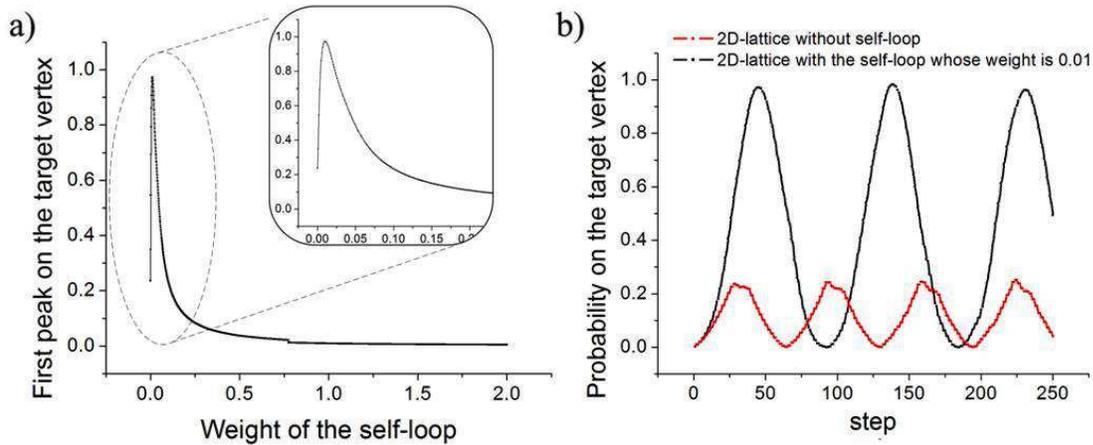}\\
  \caption{The effect of the proposed method on DTQW based search on a $20\times20$ two-dimension lattice. a) The first peak of the probability on the target vertex when the weight of self-loop change from $0$ to $2$ continuously. b) The probability on the target vertex when the weight of self-loop is $0$ and $0.01$ respectively. We notice that when the weight of the self-loop is $0$, it is equivalent to the situation where there is no self-loop on the lattice, which has been shown in Figure \ref{Complete_Mesh_loop}(b).}\label{test_loop}
\end{figure}

Firstly, let us explore the situations which cannot be tested before, i.e. the situations where the ``number'' of self-loop is not an integer. We use DTQW based search on a $20\times20$ two-dimension lattice with one self-loop on each vertex for example, but replace the original Grover coin operator with $G_j'(n)$ when perform the DTQW based search. We use the first peak of probability on the target vertex to evaluate the success rate of the search, which has been mentioned in the section ``Background''. Then we change the parameter $n$ in operator $G_j'(n)$ from $0$ to $2$ with the increment of $0.001$, for each value of $n$ we perform the DTQW based search on the lattice and record the first peak of the probability on the target vertex, to observe how the weight of the self-loop affects the success rate. Figure \ref{test_loop}(a) records how the value of the first peak changes with the weight of the self-loop. The result shows that when the weight of the self-loop is less than 0.5, the success rate (i.e. the value of the first peak) of DTQW based search changes sharply.
We find that when $n$ is about 0.01, the success rate of the search reaches the maximum. Figure \ref{test_loop}(b) shows how the probability on the target vertex evolves with step when the weight of self-loop is 0 and 0.01 respectively. We can see that when $n=0.01$, the first peak of the probability is $97.2\%$, while when $n=0$ the first peak of the probability is only $23.6\%$. We notice that when the weight of the self-loop is $0$, it is equivalent to the situation where there is no self-loop on the lattice, which has been shown in Figure\ref{Complete_Mesh_loop}(b).

The example on the $20\times20$ two-dimension lattice shows that the success rate of DTQW based search is improved by the self-loop whose weight is between $0$ and $1$, which cannot be tested before. Then, there are two questions as follows should be considered. Firstly, if the scale of the lattice changes, how to choose a proper weight of each self-loop; Secondly, whether the proposed method affects the computational complexity of the DTQW based search.

\begin{figure}[htbp]
  \centering
  \includegraphics[width=0.9\textwidth]{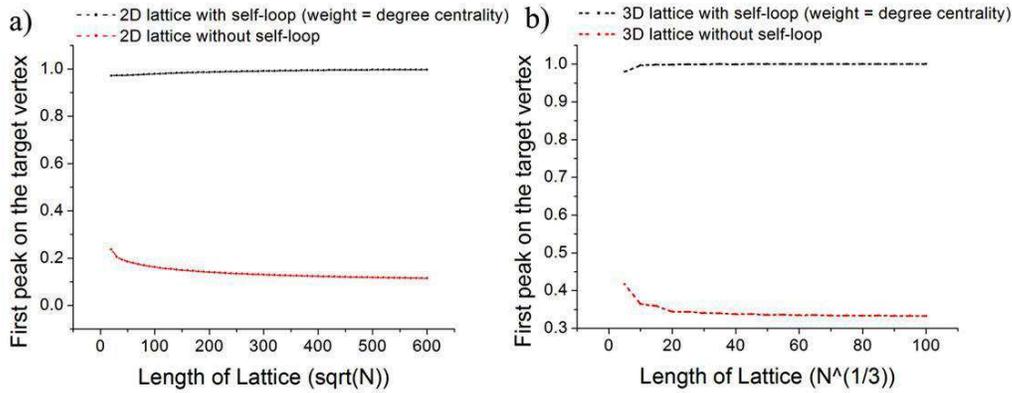}\\
  \caption{The comparison between the success rates of the original DTQW based search (on the original graphs without self-loop) and the DTQW based search enhanced by the proposed method (we attach an adjustable self-loop on each vertex, and set the weight of each self-loop to the degree centrality of the vertex). }\label{success_rate}
\end{figure}
To the first question above, we give a preliminary conclusion without proof in this paper. This preliminary conclusion is inferred from two facts as follows. Firstly, when perform DTQW based search on the complete graph, the most suitable weight of the self-loop is one\cite{Wong2015Grover}; Secondly, when perform DTQW based search on the $20\times20$ two-dimension lattice, the suitable weight of the self-loop is about $0.01$. From these two facts, we speculate that a proper weight of each self-loop may be equal to the degree centrality of its corresponding vertex, as equation (\ref{E_loop_centrality}).
\begin{equation}\label{E_loop_centrality}
  n = C_j = \frac{Deg(j)}{N-1}
\end{equation}
where $Deg(j)$ is the degree of the vertex $v_j$, and $N$ is the number of vertices on the graph. Then we verify this speculation on the two-dimension lattices and three-dimension lattices with different scale respectively, and find that it is effective on both these two types of structure. The results are shown in figure \ref{success_rate}.
From figure \ref{success_rate} we can see that the success rate of the DTQW based search algorithm enhanced by the proposed method is improved to nearly $100\%$ always, it even increases slightly with the size of the lattice. While in the original DTQW based search algorithm, the success rate is much lower, and it decreases when the scale of the lattice increase (scales as $O(1/\log{N})$ \cite{lovett2012spatial}).

\begin{figure}[htbp]
  \centering
  \includegraphics[width=0.9\textwidth]{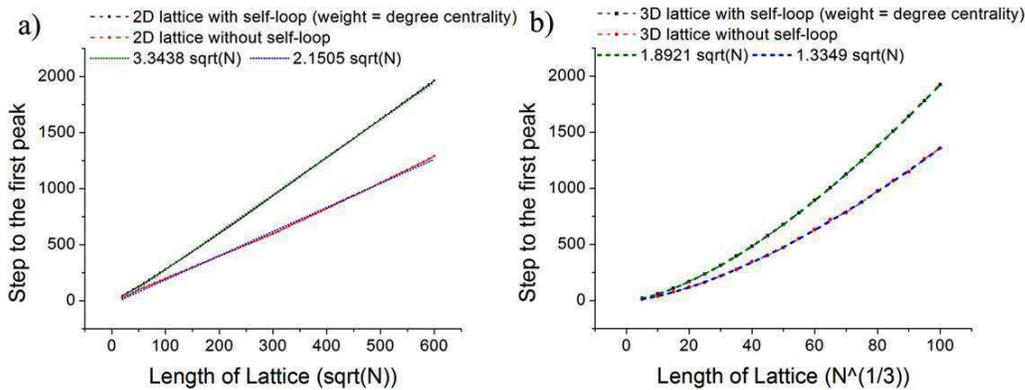}\\
  \caption{The comparison between the computational complexities of the original DTQW based search (on the original graphs without self-loop) and the DTQW based search enhanced by the proposed method (we attach an adjustable self-loop on each vertex, and set the weight of each self-loop to the degree centrality of the vertex).}\label{comlexity}
\end{figure}

To the second question, we record the step $t_{peak}$ when the first peak occurs on the target vertex on two-dimension lattices and three-dimension lattices with different scale respectively. And then analyse how $t_{peak}$ vary with the scale of the lattice. The results are shown is figure \ref{comlexity}. From the results we can see that although the proposed method makes the algorithm take a longer time to finish the search, the computational complexity of the DTQW based search algorithm is still $O(\sqrt{N})$ on two-dimension lattices and three-dimension lattices.

In this section, we successfully enhanced the success rate of DTQW based search algorithm on two-dimension lattices and three-dimension lattices by the proposed method. More specifically, when perform the DTQW based search, we attach one self-loop on each vertex and make the weight of each self-loop be equal to the degree centrality of the corresponding vertex by the proposed method. For the degree centrality is always less than one in two-dimension lattices and three-dimension lattices, this enhancement on DTQW based search cannot be realised by the original model that adjusts the weight of self-loops by the number of them. But there is no evidence that setting the weight of self-loop to the degree centrality is optimal, or it will be effective on other types of structure. For the evolutions of DTQW based search are complex to be analyzed, and it is not the topic of this paper, we do not discuss this issue further here.

\section{Conclusions}

In this paper, we propose a model of adjustable self-loop on discrete-time quantum walk. The proposed method adjusts the weight of the self-loop on discrete-time quantum walk (DTQW) by a real parameter in a specially designed coin operator, which named Grover-loop operator. Compared with the original model that adjusts the effect of self-loops by their number, the proposed method has two advantages as follows. Firstly, the new method not only generalises the situations where the number of self-loops is an integer, but also provides a way to adjust the weight of the self-loop continuously. Secondly, through the proposed method, the adjustment of the self-loop's weight do not make the dimension of walker's state space change.

Through the proposed method, the potential of DTQW in the application will be enhanced. In this paper, we improve the DTQW based search on two-dimension lattices and three-dimension lattices for example. In the original DTQW based search on two-dimension lattices, the success rate only scales as $O(1/\log N)$\cite{lovett2012spatial}, where $N$ is the size of lattices. And the success rate cannot be improved by the method that just attach self-loops on each vertex, although this method successfully improves the DTQW based search on complete graphs\cite{Wong2015Grover}. For instance, on a $20\times20$ two-dimension lattices, the success rate of DTQW based search is only $23.6\%$, and this success rate will decrease with the size of the lattice. The additional self-loops on the lattice will make the success rate even lower. However, if we attach the proposed adjustable self-loop on each vertex, and adjust the weight of each self-loop to the degree centrality of the vertex, the success rate of DTQW based search on the $20\times20$ two-dimension lattice will be boosted to $97.2\%$, and this success rate is even increase slightly with the size of the lattice. These results on two-dimension lattices also can be generalised to three-dimension lattices. To the best of our knowledge, this is the first time that such an improvement is achieved on the DTQW based search on two-dimension lattices and three-dimension lattices.

The proposed method paves a new way to explore how the self-loops affect DTQW. There are many interesting issues for further studies. Such as the influence of the adjustable self-loop with non-integer weight on DTQW on various structures, and the new potential of DTQW with the adjustable self-loop in various applications.

\section*{Acknowledgments}

This work was supported by the National Natural Science Foundation of China (NSFC) No.61221491 and the Open Fund from HPCL No. 201401¨C01.

\end{document}